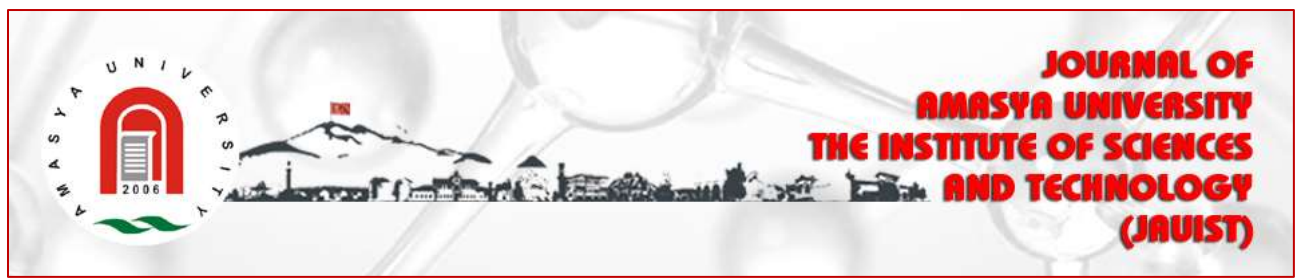

# BIBLIOMETRIC ANALYSIS OF THE WORLD SCIENTIFIC PRODUCTION IN CHEMICAL ENGINEERING DURING 2000-2011. PART 2: ANALYSIS OF THE 1,000 MOST CITED PUBLICATIONS

Ruben Miranda (✉)[1], Esther García-Carpintero[2]

[1]Departamento de Ingeniería Química y de Materiales, Facultad de CC. Químicas, Universidad Complutense de Madrid. Avda. Complutense s/n, 28040 Madrid-Spain rmiranda@ucm.es.
[2]Agencia de Evaluación de Tecnologías Sanitarias, Instituto de Salud Carlos III. Avda. Monforte de Lemos 5, 28029 Madrid-Spain eegarcia@isciii.es

**ABSTRACT**

A comprehensive bibliometric analysis of the scientific production of Chemical Engineering area has been carried out using the Web of Science database for the period 2000-2011 through three complementary studies. Part 2 demonstrated a displacement of the most cited publications to the Far East, especially due to China, however, this displacement is less important to that observed for total scientific production (Part 1). United States is still the country with the highest number of articles among the 1,000 most cited (31.5%), largely above what expected from their number of publications, followed by Germany (8.4%) and China (7.5%). The international collaboration, at least globally, seems not being an important issue for producing highly cited papers. In fact, only two from the top 25 most cited papers were international collaborations (8%). Furthermore, a large share of reviews among the 1,000 most cited papers (65%) has been observed. Although the number of institutions with more publications among the most cited in the area are from United States, the two institutions with the highest cited papers are CNRS (France) and CSIC (Spain). The most cited papers are highly concentrated in a few journals: around half of the most cited papers were published in five journals. Generally, the most cited papers are published in journals with high impact factors, however, there is also a significant number of highly cited papers published in journals with low or not having impact factor.

**ÖZET**

Kimya Mühendisliği alanının bilimsel üretiminin kapsamlı bir bibliyometrik analizi, tamamlayıcı üç çalışma aracılığıyla 2000-2011 dönemi için Web of Science veritabanı kullanılarak gerçekleştirilmiştir. Bölüm 2, özellikle Çin nedeniyle en çok alıntı yapılan yayınların Uzak Doğu'ya kaydırıldığını göstermiştir, ancak bu yer değiştirme, toplam bilimsel üretim için gözlemlenenden daha az önemlidir (Bölüm 1). Amerika Birleşik Devletleri, en çok alıntı yapılan 1000 makale (% 31,5) arasında, yayın sayısından büyük ölçüde beklenenin üzerinde, en yüksek makale sayısına sahip ülkedir, onu Almanya (% 8,4) ve Çin (% 7,5) izlemektedir. Uluslararası işbirliği, en azından küresel olarak, çok alıntılanan makaleler üretmek için önemli bir konu değil gibi görünmektedir. Aslında, en çok alıntı yapılan 25 makaleden sadece ikisi uluslararası işbirlikleri (% 8) ile yapılmıştır. Ayrıca, en çok alıntı yapılan 1.000 makale arasında yorumların büyük bir kısmı (% 65) gözlemlenmiştir. Alanında en çok atıfta bulunulan kurumlar arasında daha fazla yayını olan kurumların sayısı Amerika Birleşik Devletleri'nden gelse de, en çok atıf alan iki kurum CNRS (Fransa) ve CSIC (İspanya)' dır. En çok alıntı yapılan makaleler birkaç dergide yoğunlaşmıştır: En çok alıntı yapılan makalelerin yaklaşık yarısı beş dergide yayınlanmıştır. Genellikle, en çok atıf alan makaleler yüksek etki faktörlerine sahip dergilerde yayınlanmakta, ancak etki faktörü düşük olan veya olmayan dergilerde yayınlanan önemli sayıda yüksek atıf alan makaleler de vardır.

**Anahtar Kelimeler:** Kimya mühendisliği; en çok alıntı yapılan makaleler; bilimsel dergiler; etki faktörü

## 1. INTRODUCTION

Although citation rate is not a direct measure of the impact or importance of a particular work, it provides a marker of its recognition within the scientific community and frequently, the best manuscript can be considered the one most cited in peer-reviewed journals (Ho, 2012). In fact, citation analysis has been accepted as a popular method for measuring the impact of an article, a researcher, a country or region, being the Journal of Citation Reports (JCR) from Web of Science the most widely used tool for this type of analysis.

As commented in Part 1, most of the previous bibliometric studies in Chemical Engineering focused only in a specific geographical region or country (Yin, 2009; Fu et al., 2014; García-Carpintero and Miranda, 2013; Rojas-Sola and de San-Antonio-Gómez, 2010a, 2010b), certain journals, i.e. Chemical Engineering Journal and Biochemical Engineering Journal (A. Shubert, 1998) or researchers of specific research institutes, i.e. the Institute of Chemical Engineering in Taiwan (Chang and Cheng, 2012).

Furthermore, the bibliometric studies of highly cited papers are also rather limited in this area. To the best knowledge of the authors, there are only two bibliometric studies of the most cited papers of the area: Ho (2012) and Chuang et al. (2013). First, Ho (2012) focused on the articles in Chemical Engineering area with more than 100 cites since 1931 to 2010 using the Science Citation Index from Web of Science, which turns into a sample of 3,828 articles published between 1935 and 2001. Around 4% of these articles were published before 1960, a 9% in the 1960s, 16% in 1970s, 24% in 1980s, 31% in 1990s and 16% in 2000s. On the other hand, Chuang et al. (2013) focused on the use of the Essential Science Indicators (ESI) from Thomson Reuters as an alternative database for the analysis high-impact papers and Chemical Engineering area was used as example. A sample of 475 ESI papers were analyzed, published from January 1$^{st}$ 1999 to October 31$^{st}$ 2009.

In the study of Ho (2012), the analyzed period was very long (1935-2001). Most of the papers were published before 2000 (84%), and it does not represent the present situation in the area. On the other hand, the analyzed period considered by Chuang et al. (2013) was closer to present, but ESI indicators have used instead. Although a number of bibliometric indicators were calculated in these studies, the analysis were partial. Ho (2012) analyzed the publication performances by countries, institutions and authors, while Chuang et al. (2013) also analyzed the journals and the keyword titles. However, none of them analyzed the evolution of the number of high-impact papers by countries with time and they did not calculated bibliometric indicators based on the received cites (total cites, cites per article and year since publication, average impact factor of the publications by country). In addition, no collaboration networks among countries were analyzed. Furthermore, and perhaps the most important lack of these studies is that none of them compared the general trends in the area (total scientific production) and those followed by the highly cited papers. Whatever the case, present manuscript will benefit from the data obtained by these previous studies as the results obtained will be compared with the trends observed in the past, comparing with the work of Ho (2012), and with the results obtained analyzing ESI indicators for a similar period of time, using the work of Chuang et al. (2013).

Therefore, the objective of the present study is the bibliometric analysis of the 1,000 most cited papers in the area of Chemical Engineering published in 2000-2011, distinguishing between different countries and the evolution of these indicators with time. Especifically, it will be analyzed: a) the total number of publications, b) the impact factor of the journals where the most cited articles were published, c) the total number of cites and the number of cites per article and year since published, d) the journals publishing the highly cited articles, e) the degree of international collaboration in the most cited papers, and f) the research organizations producing the top cited papers.

## 2. METHODOLOGY

Scientific production in the area of Chemical Engineering, was obtained from Web of Science® database. WoS is a highly reputed on-line service for scientific information, developed by the *Institute for Scientific Information (ISI)* and now part of Clarivate Analytics®. WoS publishes the *Journal of Citation Report (JCR)*, one of the most prestigious databases for cites analysis of the publications.

Present study has been circumscribed to the journals indexed in the area of "Engineering, Chemical" of the WoS, the number of these journals varying from 110 to 135 during the analyzed period (2000-2011). The analyzed publications include the articles and reviews which have been published in the journals which were indexed in the Chemical Engineering area according to JCR 2011, although these journals were not previously indexed in this area. In such way, data along the analyzed period can be compared without being influenced by the number of journals indexed in this area with time. Although the journals indexed in the thematic area of "Engineering, Chemical" do not represent 100% of the scientific production in Chemical Engineering, they represent a good sample of the research carried out in this area. In these sense, most of the previous works carried out in the area followed this approach (Rojas-Sola and San Antonio Gómez, 2010a, 2010b; Ho, 2012; Chuang et al., 2013; García-Carpintero and Miranda, 2013; Fu et al., 2014). Furthermore, JCR also allows one journal to be indexed in several thematic areas, which facilitates to be taken into account some thematic areas related to the main thematic area.

The total citations of the articles published in the period 2000-2011 were evaluated at mid-August 2012 (TC2012), and the 1,000 most cited publications were selected for their analysis. To analyze the evolution of the most cited publications with time, a different sample was used. In this case, the 100 most cited papers by publication year were considered (1,200 publications).

## 3. RESULTS AND DISCUSSION

### 3.1. Documents characteristics

The 1,000 most cited publications represented around a 0.47% of the total publications in the area in the period 2000-2011 (213,264 publications). These publications received a total of 169,227 cites, which means an average of 169.2 cites since publication to 2012. The number of citations received for the 1,000 most cited papers varied from 104 to 1,231, the cumulative distribution of cites received by these articles is shown in Figure 1a. The top 100 most cited articles received an average total citations of 410.2, which is similar to the average number of citations received for the 100 most cited articles in other areas, i.e. 300-700 in different medical fields (Hsu and Ho, 2014). On the other hand, the number of cites per article and year since publication varied largely from 8.7 to 136.7, with an average of 20.5 and a median of 16.2 cites per publication and year (see distribution curve shown in Figure 1b). For the top 25 most cited publications published in the period 2000-2011, the number of cites per year since its publication varied from 38.8 to 136.7, with an average of 66.6 and a median of 57.9.

The distribution of publications by publication year was the following: 158 (2000), 146 (2001), 148 (2002), 151 (2003), 148 (2004), 90 (2005), 75 (2006), 43 (2007), 22 (2008), 14 (2009), 4 (2010) and 1 (2011). As expected, most highly cited papers were published in the first years of the analyzed period as newly published articles require time to accumulate citations (Lefaivre et al., 2011; Hsu and Ho, 2014). Previous studies showed that highly cited articles reach its citation peak in the first several years after publication, this citation peak occurring at 3-7 years depending on the area considered (Hsu and Ho, 2014). In the present study, the number of publications among the most cited is similar in the period 2000-2004 and starts decreasing since 2005. As the total number of citations was evaluated at 2012, this suggest that the citation peak of the most cited papers in Chemical Engineering could occur after than in other areas.

Most of the highly cited papers were reviews (650, 65%) and 350 were articles (35%). The reviews are clearly over-represented compared to the share of reviews in total scientific production in the area (98.5% articles and only 1.5% reviews). Review papers tend to be cited more frequently than other types (Glänzel and Moed, 2002; Moed, 2010; Ho and Kahn, 2014; Miranda and Garcia-Carpintero, 2018). Previous studies observed that reviews usually received around 3 times more citations than articles, but varying largely among research areas (Miranda and Garcia-Carpintero, 2018). However, rough estimations carried out indicated that the reviews in Chemical Engineering were cited from 4 to 5 times more the articles in the period 2000-2010. The over representation found in Chemical Engineering is considerably larger than in previous studies focused in Chemical Engineering and in other scientific areas. Chuang et al. (2013), for example, found that a 20% of the most cited papers in Chemical Engineering area were reviews in ESI database. On average, a 12% of the most cited papers in SCI (considering all disciplines) are reviews (Askness, 2003).

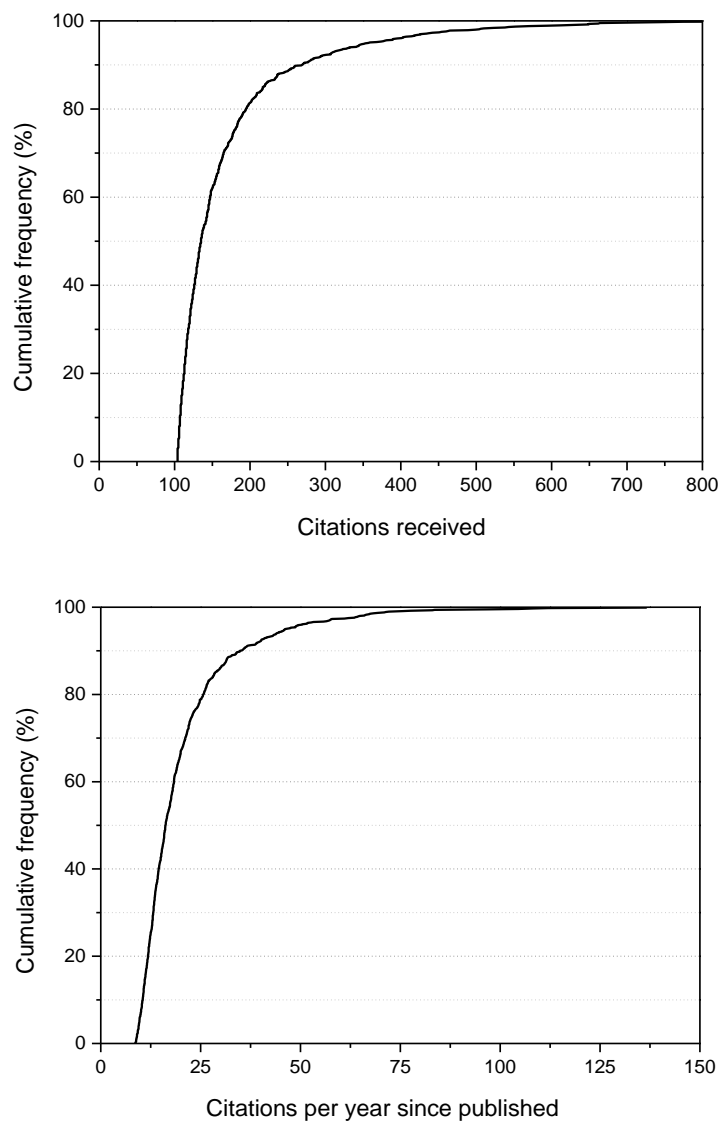

**Figure 1.** Cumulative distribution of the citations received by the 1,000 most cited papers of the period 2000-2011 in the area of Chemical Engineering: (a) average citations per paper and (b) average citations per paper and year since published.

The share of reviews in the most cited papers can increase largely in some specific areas, Ho and Kahn (2014) determined that a 47% of the highly cited reviews were from four main areas: Biochemistry and Molecular Biology, Multidisciplinary Chemistry, Multidisciplinary Sciences and Cell Biology.

### 3.2. Scientific production by geographic areas and countries

First, the distribution of the 1,000 most cited articles has been analyzed by geographical areas. The highest share in the most cited papers is from North America (33.9%), followed by Europe (33.7%) and Asia (31.3%), with almost the same rates. The share of the other three regions together (Oceania, Central and South America and Africa) was only 5.8%. If these values are compared with their shares in total number of publications in the same period (Figure 2), it is observed that North America is largely over represented in the most cited papers (33.9%) compared to its share in total publications (19.3%). Europe was represented in the 1,000 most cited papers as it could be expected by its total number of publications, 33.7% vs. 34.6%, similarly to what occurred for Oceania. On the contrary, Asia is under represented (31.3% in the most cited papers compared to 40.5% total publications) and this also happened for Central and South America, and especially for Africa.

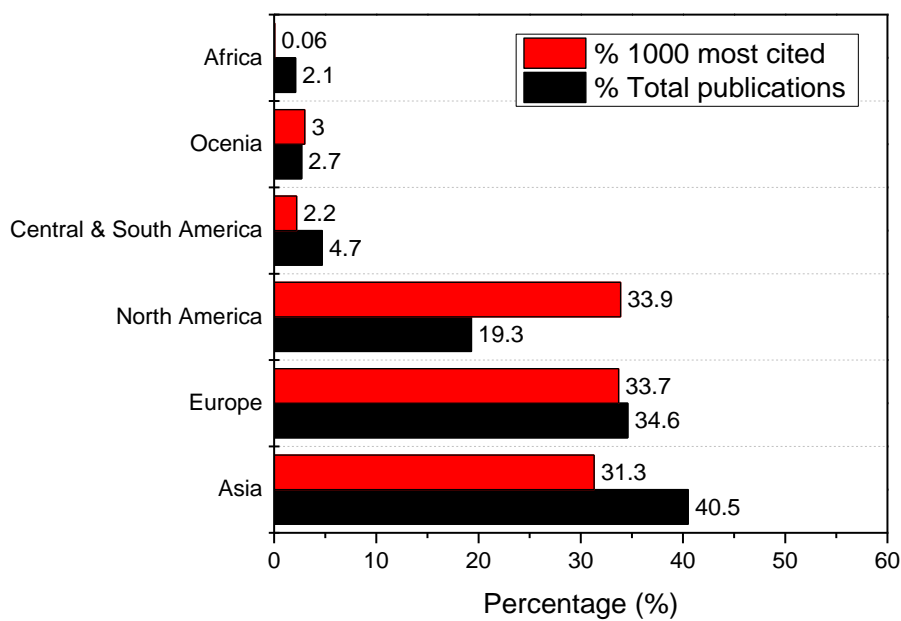

**Figure 2.** Comparison of the % of total publications and % among the 1,000 most cited publications by geographical regions in the area of Chemical Engineering during 2000-2011.

North America participated in 339 of the 1,000 most cited publications: United States in 315 and Canada in 24. A total of 24 different European countries participated in 337 of the 1,000 most cited publications, the most productive countries being Germany (84), France (65), Spain (62), United Kingdom (58), Italy (42), Netherlands (35), Switzerland (24), Greece (22) and Denmark (20). Thirteen different Asian countries participated in 313 publications amongst the most cited, the most productive countries being: China (75), India (65), Japan (60), Turkey (41), South Korea (20), Taiwan (15). Oceania participated in 30 publications by two different countries: Australia (26) and New Zealand (4). Central and South American countries participated in 22 of the 1,000 most cited publications: Argentina (7), Brazil (5), Mexico (4), Chile (3), Venezuela (2) and Colombia (1), and finally, two countries from Africa participated in 6 publications of the 1,000 most cited: Tunisia (5) and South Africa (1).

A total of 51 different countries participated in the 1,000 most cited papers, and 25 different countries participated in the top 100 most cited papers of the period: 9 countries with one publication and 16 with two or more publications. The participation of different countries in the most cited papers is significantly higher than in other research areas. In fact, the highly cited articles in different areas are usually originated from only a small number of countries. For example, the number of countries participating in the top 100 most cited articles were 10 in the area of health care and medical services, 10 in ophthalmology, 9 in anesthetic journals and 6 countries in general surgery (Hsu and Ho, 2014).

A large difference between the United States and the other countries was clearly observed (Figure 3). A 31.5% of the most cited articles have at least one author from the United States. Although United States is also the country with the highest number of publications during the analyzed period (32,597; 15.3% total publications), the share of papers among the 1,000 most cited in the area is around double than its share in total publications. The prevalence of the United States was even higher in the past. When the articles with more than 100 citations in the area of Chemical Engineering in the Science Citation Index during the period 1931-2010 (a 54% of these articles were published before 1991) was analyzed, the percentage of highly cited papers from United States was 49% (Ho, 2012).

An overwhelmingly predominance of the United States is still clear in the top cited papers in all the scientific areas, its predominance being even more important for top-1% cited publications compared to top-10% cited publications (Leydesdorff et al., 2014). Although this prevalence is decreasing continuously with time, United States is still over represented in the most cited papers of virtually almost all scientific areas. For example, United States participated in 37% of the highly cited papers in the water resources area (Chuang et al., 2011), in the 48% of the highly cited papers in materials science (Ho, 2014) and in the 49% of the highly cited articles on biomass (Chen and Ho, 2015).

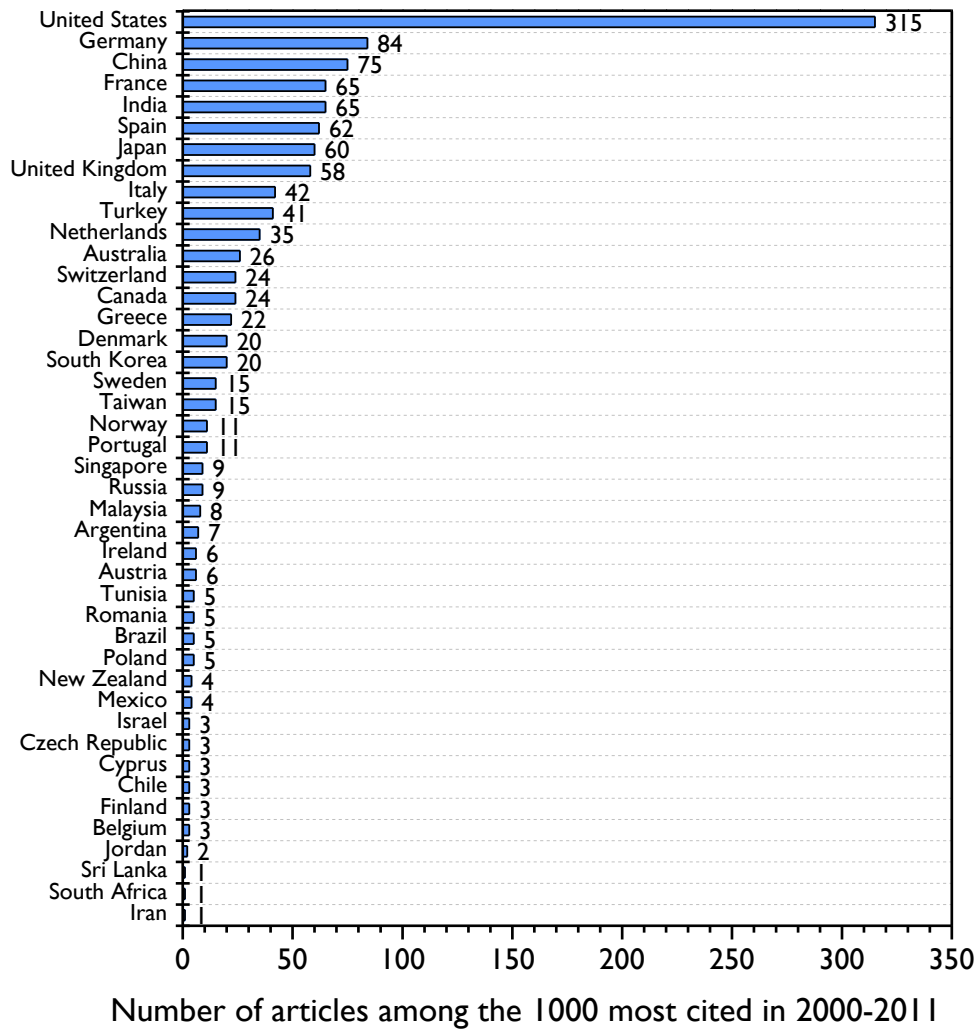

**Figure 3.** Number of articles among the 1,000 most cited by nationalities during the period 2000-2011.

The second and third countries with more articles among the most cited were Germany and China, participating in an 8.4% and 7.5% of total articles, respectively. They were followed by India, France, Spain, Japan and United Kingdom, participating each around 6% of the most cited papers of the analyzed period. Finally, Italy, Turkey and the Netherlands, participated in around 4% each of the most cited papers. It is important to notice that China had traditionally published in journals having lower citation impact than the world average and received lower citation rates than expected according to its number of publications (Yie and Jie, 2012). In Chemical Engineering area, China was ranked the 2$^{nd}$ country in number of publications in the period 2000-2011 (26,110 publications, 12.24%), and participated only in 7.5% of the most cited papers (3$^{rd}$ ranked).

There are also countries which are not among the most productive countries but participating in higher number of publications among the 1,000 most cited than them. This is the case of Denmark (16$^{th}$ ranked in top cited papers with 20 articles, but 35$^{th}$ ranked in terms of total publications), Norway (20$^{th}$ ranked in top cited papers with 11 articles, but 36$^{th}$ ranked in total publications), Malaysia (24$^{th}$ ranked in top cited papers with 8 articles but 31$^{st}$ ranked in total publications), Austria (26-27$^{th}$ ranked in top cited articles with 6 articles, 39$^{th}$ ranked in total publications), Ireland (26-27$^{th}$ ranked in top cited articles with 6 articles but 46$^{th}$ ranked in total publications) and Tunisia (28-30$^{th}$ ranked in top cited articles with 5 publications but 49$^{th}$ ranked in total publications).

For comparative purposes, the percentage of articles among the 1,000 most cited compared to the total scientific production of each country was calculated, both for the 30 most productive countries and the 30 countries with highest number of publications among the 1,000 most cited. The results are shown in Figure 4. Without considering Sri Lanka and Cyprus, with very low scientific production to be significant, the average number of publications among the 1,000 most cited varied from 0.02% (Iran) to 1.68% (Denmark) of their total number of publications.

Denmark clearly outstands from the other countries (1.68%), followed by Switzerland (1.20%), Greece (1.07%), Ireland (0.97%) and United States (0.95%). As demonstrated by these results, although the ratio of top cited papers among the total scientific production of the United States is one of the highest, there are countries with lower scientific production but more successful in producing highly cited papers. On the opposite, Iran (0.02%), Poland and South Africa (0.08%), Brazil (0.12%), Romania (0.15%) and Russia (0.16%) are those countries with the lowest scientific production among the most cited. As expected, these results are in agreement with the quality indicators (total cites, impact factor, etc.) analyzed in Part 1 of this article for the different countries.

As recognized by Leydesdorff et al. (2014), the share of the major countries of Europe (United Kingdom, Germany and France) in the highly cited papers for all the areas in WoS (top 1% and top 10%) is decreasing, probably by the rapid increase of countries such as China and other emerging countries. However, Italy and the Netherlands, had a marginally upward slope in the share of the highly cited publications in different scientific areas, which is similar to what observed in the present study for the Chemical Engineering area. On the contrary, Spain, Italy or Greece performed above expectation in Chemical Engineering while, considering all the scientific areas, Spain and Italy performed almost as expected and Greece remained below expectation (Leydesdorff et al., 2014).

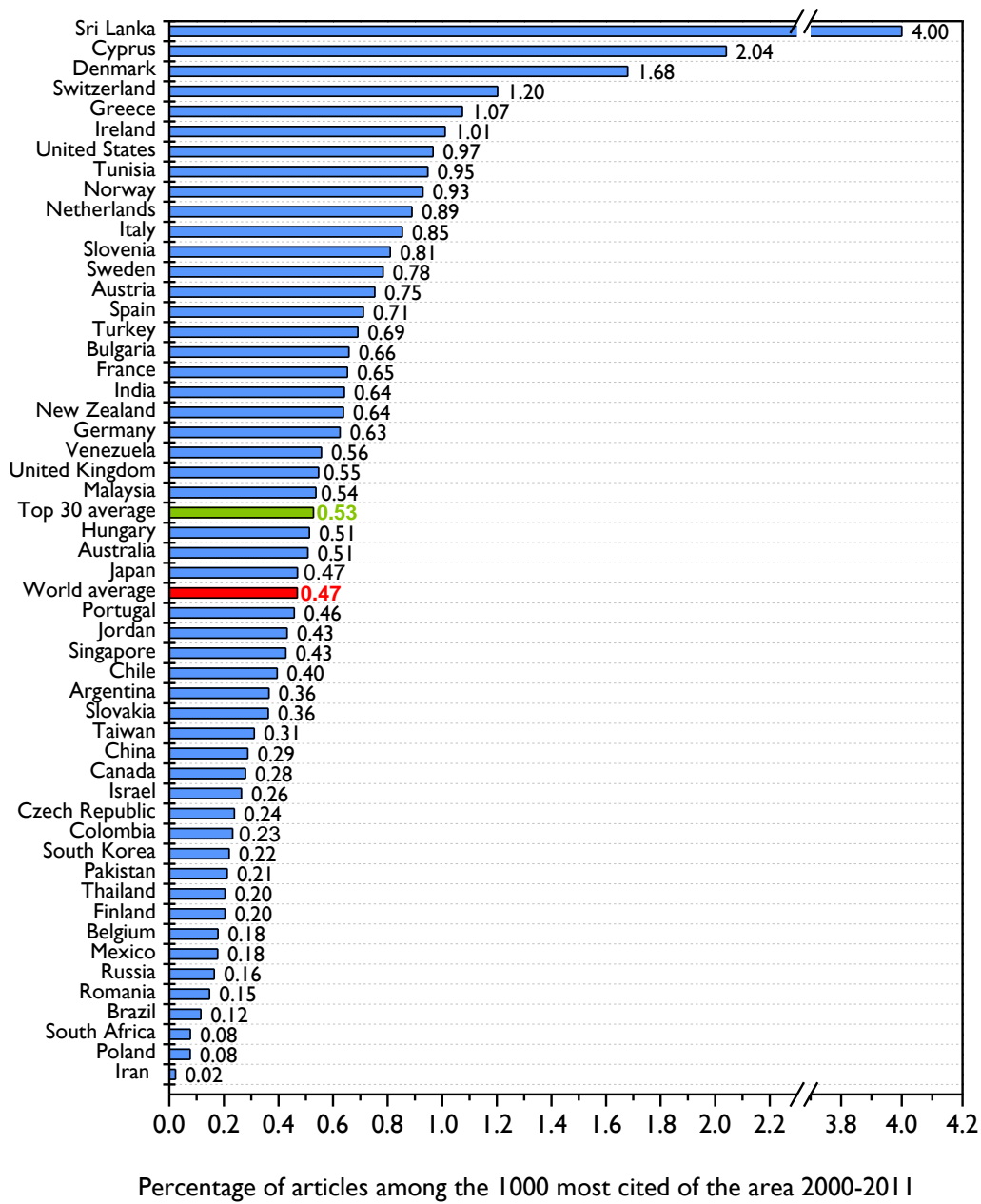

**Figure 4.** Percentage of publications among the 1,000 most cited of the area from total publications during the period 2000-2011 by countries with publications among the 1,000 most cited.

To analyze the evolution with time of the nationalities of the authors of the most cited papers, the 100 most cited articles per publication year were analyzed (1,200 total documents). Table 1 shows the results obtained. The importance of United States maintained almost constant, and only decreased slightly despite the share of United States in total publications decreased largely from 21.1% in 2000 to 13.3% in 2011. Although the predominance of United States is still clear (366 publications, 30.5%), a clear increase of the importance of China and Turkey was observed in the last years. Considering the 1,000 most cited papers in 2000-2011, China was ranked $3^{th}$ (7.5% of the 1,000 most cited papers) but considering the 100 most cited papers in each year, China was ranked $2^{th}$ (9.4%). The improvement of Turkey is even higher: it was ranked $10^{th}$ in the 1,000 most cited papers of the period 2000-2011 (4.1% of the 1,000 most cited papers) but ranked $3^{rd}$ regarding the 100 most cited papers in each year (7.0%). A large increase in the importance of Turkey during the analyzed period, especially since 2009 was observed. In fact, Turkey became the country with the highest number of publications among the 100 most cited articles in 2011 (33 publications vs. 29 from United States and 12 from China).

There are some peaks in some countries at certain publication years, i.e. for India in 2004-05 or Spain in 2006-08, however, the average contribution of these countries to the most cited papers is similar when considering the 1,000 most cited articles in the period 2000-2011 or the number of articles among the 100 most cited paper in each publication year: 6.5% vs. 5.8% for India and 6.2% vs. 5.8% for Spain. A similar case was observed for United Kingdom, with an average of 5.8% of the 1,000 most cited publications and 5.7% if the 100 most cited papers of the year are considered.

**Table 1.** Articles among the 100 most cited per publication year by countries during the period 2000-2011.

| Rank | | 2000 | 2001 | 2002 | 2003 | 2004 | 2005 | 2006 | 2007 | 2008 | 2009 | 2010 | 2011 | 2000-2011 | % 2000-2011 |
|---|---|---|---|---|---|---|---|---|---|---|---|---|---|---|---|
| 1 | United States | 34 | 32 | 29 | 33 | 29 | 39 | 35 | 26 | 27 | 30 | 23 | 29 | 366 | 30.5 |
| 2 | China | 9 | 5 | 5 | 7 | 6 | 8 | 13 | 11 | 5 | 15 | 17 | 12 | 113 | 9.4 |
| 3 | Turkey | 4 | 4 | 5 | 1 | 5 | 4 | 1 | 4 | 2 | 8 | 13 | 33 | 84 | 7.0 |
| 4 | Germany | 9 | 9 | 8 | 6 | 6 | 7 | 4 | 4 | 8 | 4 | 8 | 4 | 77 | 6.4 |
| 5 | India | 6 | 7 | 3 | 7 | 13 | 10 | 4 | 3 | 4 | 5 | 6 | 1 | 69 | 5.8 |
| 6 | Spain | 6 | 3 | 5 | 5 | 7 | 3 | 8 | 13 | 9 | 7 | 2 | 1 | 69 | 5.8 |
| 7 | United Kingdom | 5 | 3 | 9 | 6 | 5 | 9 | 1 | 5 | 11 | 7 | 5 | 2 | 68 | 5.7 |
| 8 | Japan | 7 | 11 | 5 | 7 | 5 | 2 | 7 | 1 | 4 | 3 | 3 | 2 | 57 | 4.8 |
| 9 | France | 5 | 9 | 9 | 3 | 7 | 6 | 5 | 4 | 2 | 2 | 3 | 2 | 57 | 4.8 |
| 10 | Netherlands | 8 | 5 | 2 | 5 | 0 | 4 | 5 | 3 | 4 | 5 | 4 | 3 | 48 | 4.0 |
| 11 | Canada | 1 | 3 | 1 | 2 | 4 | 2 | 7 | 2 | 6 | 6 | 3 | 4 | 41 | 3.4 |
| 12 | Italy | 4 | 4 | 3 | 2 | 3 | 1 | 4 | 3 | 4 | 8 | 2 | 3 | 41 | 3.4 |
| 13 | South Korea | 3 | 0 | 2 | 3 | 6 | 4 | 3 | 2 | 2 | 6 | 1 | 4 | 36 | 3.0 |
| 14 | Australia | 3 | 3 | 1 | 1 | 0 | 3 | 4 | 5 | 6 | 4 | 2 | 0 | 32 | 2.7 |
| 15 | Denmark | 2 | 4 | 3 | 2 | 2 | 0 | 1 | 2 | 3 | 0 | 5 | 4 | 28 | 2.3 |
| 16 | Switzerland | 2 | 2 | 2 | 3 | 2 | 4 | 2 | 1 | 4 | 1 | 0 | 4 | 27 | 2.3 |
| 17 | Singapore | 1 | 0 | 0 | 2 | 0 | 2 | 1 | 2 | 3 | 2 | 4 | 6 | 23 | 1.9 |
| 18 | Greece | 0 | 0 | 4 | 6 | 4 | 1 | 2 | 1 | 2 | 0 | 0 | 0 | 20 | 1.7 |
| 19 | Portugal | 0 | 0 | 1 | 1 | 4 | 1 | 2 | 3 | 3 | 1 | 2 | 0 | 18 | 1.5 |
| 20 | Malaysia | 1 | 0 | 1 | 0 | 2 | 1 | 0 | 0 | 2 | 4 | 4 | 2 | 17 | 1.4 |
| 21 | Sweden | 1 | 2 | 2 | 1 | 1 | 0 | 0 | 3 | 2 | 1 | 1 | 0 | 14 | 1.2 |
| 22 | Taiwan | 0 | 2 | 2 | 0 | 2 | 2 | 2 | 2 | 0 | 0 | 0 | 0 | 12 | 1.0 |
| 23 | Poland | 0 | 2 | 0 | 3 | 0 | 0 | 0 | 3 | 1 | 0 | 1 | 1 | 11 | 0.9 |
| 24 | Iran | 0 | 0 | 0 | 1 | 0 | 0 | 0 | 1 | 3 | 1 | 1 | 4 | 11 | 0.9 |
| 25 | Norway | 1 | 0 | 1 | 3 | 2 | 0 | 0 | 2 | 1 | 0 | 1 | 0 | 11 | 0.9 |
| 26 | Tunisia | 0 | 0 | 2 | 0 | 1 | 0 | 1 | 5 | 0 | 0 | 0 | 0 | 9 | 0.8 |
| 27 | Ireland | 0 | 0 | 0 | 1 | 1 | 0 | 0 | 0 | 1 | 2 | 2 | 1 | 8 | 0.7 |
| 28 | Brazil | 5 | 0 | 1 | 0 | 0 | 0 | 1 | 0 | 0 | 0 | 0 | 0 | 7 | 0.6 |
| 29 | Russia | 1 | 1 | 1 | 0 | 0 | 0 | 1 | 0 | 1 | 0 | 0 | 1 | 6 | 0.5 |
| 30 | Argentina | 2 | 0 | 0 | 0 | 1 | 1 | 1 | 1 | 0 | 0 | 0 | 0 | 6 | 0.5 |
| 31 | Austria | 2 | 0 | 1 | 0 | 0 | 0 | 0 | 0 | 1 | 1 | 1 | 0 | 6 | 0.5 |
| 32 | Mexico | 0 | 0 | 0 | 1 | 0 | 0 | 1 | 1 | 0 | 2 | 0 | 0 | 5 | 0.4 |
| 33 | Finland | 0 | 0 | 0 | 0 | 1 | 0 | 1 | 0 | 2 | 0 | 1 | 0 | 5 | 0.4 |
| 34 | Belgium | 0 | 0 | 0 | 0 | 0 | 0 | 0 | 0 | 3 | 0 | 0 | 1 | 4 | 0.3 |
| 35 | Romania | 0 | 0 | 0 | 0 | 0 | 0 | 0 | 1 | 0 | 0 | 0 | 0 | 1 | 0.1 |
| 36 | Egypt | 0 | 0 | 0 | 0 | 0 | 0 | 0 | 0 | 1 | 0 | 0 | 0 | 1 | 0.1 |

On the other hand, the importance of Japan and France clearly decreased during the analyzed period. France was ranked fourth in the 1,000 most cited papers (6.4% total publications) and Japan, 7th (6.0% total publications). At the beginning of the period these countries published 15% each of the 100 most cited articles by year, however, at the end of the period, they published only 2% each. Both countries published 57 documents among the 100 most cited in each publication year, i.e. 4.8%. The importance of Germany also decreased slightly. It was ranked the 2nd among the 1,000 most cited papers (8.4%) but is ranked 4th (6.4%) considering the 100 most cited paper of each publication year.

The importance of Iran has increased significantly with time, but still very far to its share of total publications, while it was the 29th country in terms of total publications in 2000, Iran is already the 4th country in 2015, only behind United States, China and India. While only one paper from Iran was amongst the 1,000 most cited papers in 2000-2011 (0.1%, ranked 43th), this value increased to 11 publications (1.0%, 24th ranked) when the 100 most cited papers of each publication year was considered. Iran even achieved a 4.0% of the 100 most cited articles published in 2011. It seems Iran was following the same pattern than China, first an increase in the production rates is observed, and then, an increase in the quality of their publications.

It is also interesting to notice that, although the papers from the United States and the European Union are becoming similar in terms of citations (Part 1 of this study), they are still far apart in the most cited papers, as also observed previously in the top segment of the 1% most highly cited papers of all the scientific areas covered by WoS (Leydesdorff and Wagner, 2009).

During the analyzed period (2000-2011), a clear displacement of the scientific production in Chemical Engineering to the Far East was observed, mainly due to China, but also to countries such as India and Iran (Part 1 of this study). When the same analysis was carried out for the 100 most cited articles by publication year, only a slight displacement to the Far East was observed (see Figure 5). The main reason is that United States maintained the highest performance. Turkey has largely increased its importance but its geographical location is very close to the gravity center and the importance of Iran in the most cited papers is still minor compared to the increase in total publications.

The plot representing the "gravity center" of the most cited publications in Chemical Engineering and its evolution with time shown in Figure 5 was calculated considering the number of publications of the countries together with their geographical coordinates. The latitude and altitude of this "gravity center" have been calculated for the different years as follows:

$$(\text{Latitude}, \text{Altitude}) = \left( \frac{\sum_{i=1}^{j} N_i \cdot \text{Latitude}_i}{\sum_{i=1}^{j} N_i}, \frac{\sum_{i=1}^{j} N_i \cdot \text{Longitude}_i}{\sum_{i=1}^{j} N_i} \right)$$

where $N_i$ is the number of publications of each country i, and $\text{Latitude}_i$ and $\text{Longitude}_i$ are the geographical coordinates of the capital of each country.

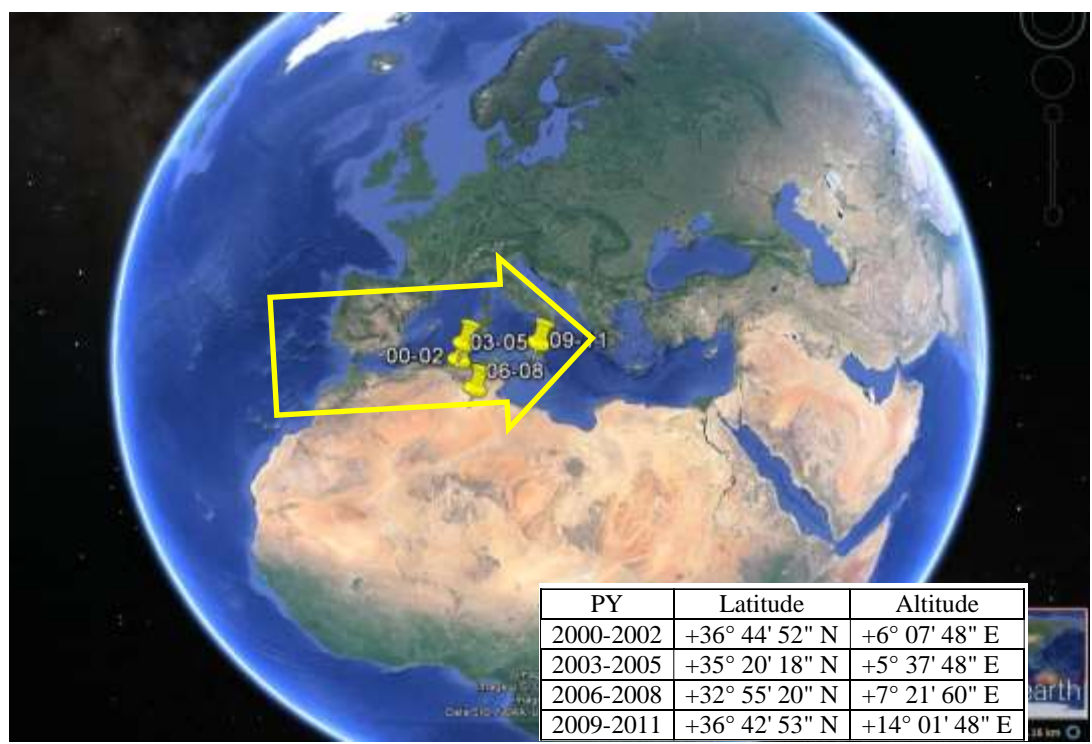

| PY | Latitude | Altitude |
|---|---|---|
| 2000-2002 | +36° 44' 52" N | +6° 07' 48" E |
| 2003-2005 | +35° 20' 18" N | +5° 37' 48" E |
| 2006-2008 | +32° 55' 20" N | +7° 21' 60" E |
| 2009-2011 | +36° 42' 53" N | +14° 01' 48" E |

**Figure 5.** Evolution with time of the gravity center of the 100 most cited articles per year in different periods (2000-02, 2003-05, 2007-08, 2009-11) in Chemical Engineering.

### 3.3. Impact factor of the journals where the scientific production was published

The average and median impact factor of the 1,000 most cited papers was 2.325 and 1.933, respectively. As the number publications among the 1,000 most cited by publication year is different and there are more highly cited articles at the beginning than at the end of the period, yearly averages were considered. In this case, the average yearly impact factor in the period 2000-2011 was 5.94, varying from 1.76 in 2000 to 31.7 in 2011. The average yearly median impact factor of the 1,000 most cited papers was 5.23, varying from 1.59 in 2000 to 31.7 in 2011. These results are largely affected by the yearly impact factors of 2010 and 2011, where only a few number of articles were published (4 in 2010 and 1 in 2011). Excluding these data, the annual average impact factor varied from 1.76 to 5.20 (average 2.81) and the median from 1.59 to 5.25 (average 2.43) in the period 2000-2009. If total scientific production is considered, average yearly impact factor varied from 0.71 in 2000 to 2.51 in 2011 (average 1.37) and the median yearly impact factor varied from 0.61 in 2000 to 2.17 in 2011 (average 1.29). These results are summarized in Figure 6.

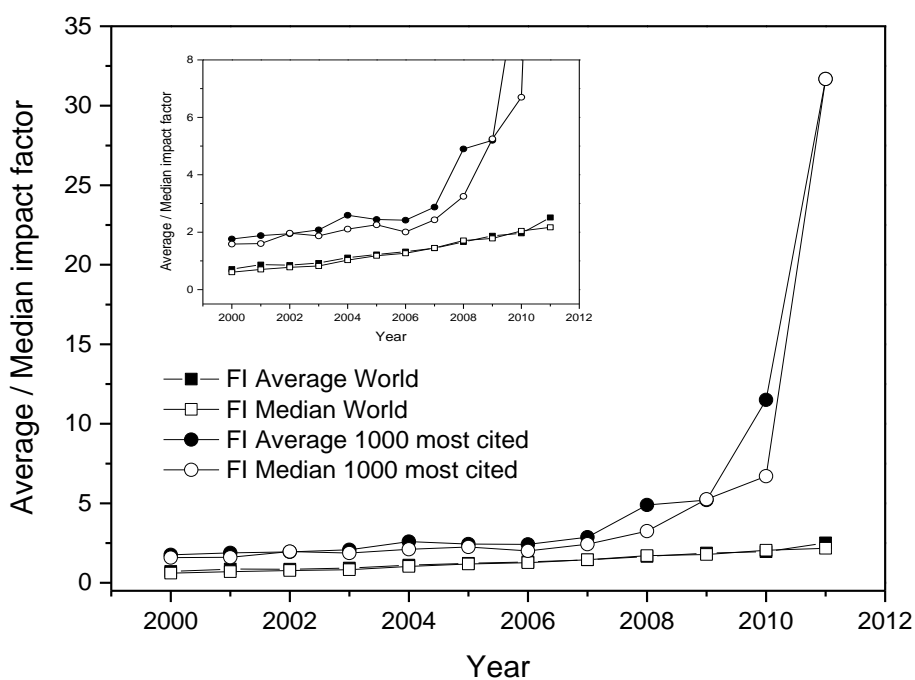

**Figure 6.** Evolution with time of the average and median impact factor for both, total number of publications and the 1,000 most cited papers in the area of Chemical Engineering in the period 2000-2011.

Figure 7 shows a box-diagram representing the minimum value, the 25$^{th}$ percentile, the 50$^{th}$ percentile, the average, the 75$^{th}$ percentile, the 99$^{th}$ percentile and the maximum value for the impact factor of the publications by year. As Figure 7 shows, there is a large variation in the impact factor of the journals where the top cited articles were published. Although the most cited articles are mainly published in high impact journals, there is also a significant amount of articles published in journals with low or not having impact factors. In the analyzed period, for example, a 13% of the 1,000 most cited publications were published in journals with impact factors lower than 1. Similar findings were obtained previously in the literature. Highly cited papers are not only published in high impact journals of a high prestige as it could be expected (Ho, 2007; Chuang and Ho, 2014) but also published in journals which even did not have impact factors, as it was observed for Materials Science (Ho, 2014), in the area of adsorption in environmental science (Ho, 2007) or in surgery area (Paladugu et al., 2002). In the biomass area, for example, 10% of the highly cited publications were published in journals that had no impact factor in JCR (Chen and Ho, 2015).

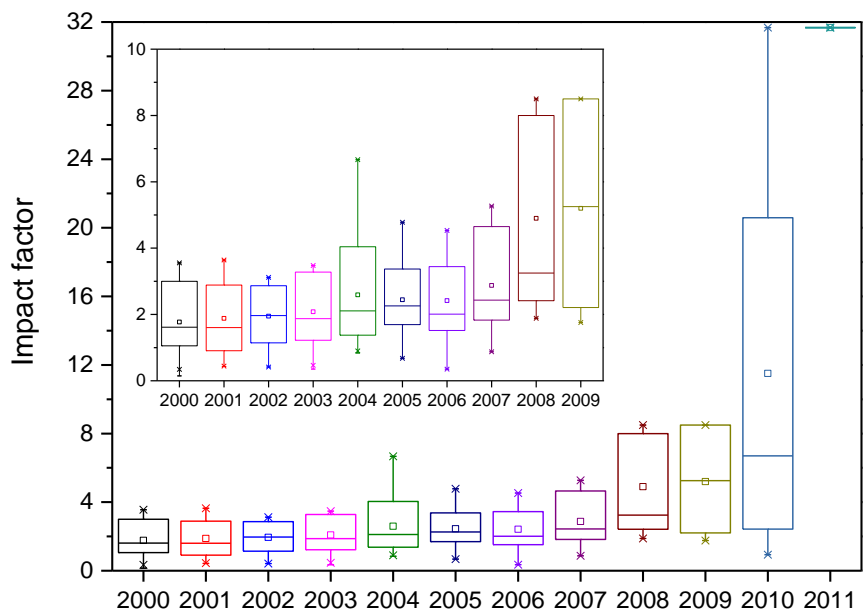

**Figure 7.** Impact factor by publication year in the area of Chemical Engineering in the period 2000-2011. The box represents the 25$^{th}$, 50$^{th}$ and 75$^{th}$ percentiles, □ is the average, and "x" represents the minimum and maximum values.

### 3.4. Total citations

During the period 2000-2011, the 213,264 articles published in the Chemical Engineering area received a total of 1,970,755 cites since publication to August 2012, meaning an average of 9.25 cites per article (Part 1 of this study). The 1,000 most cited papers received 169,227 cites, which means an average of 169.2 cites. In other words, the 1,000 most cited papers received 8.59% of total citations in the area despite they represented only 0.47% total publications.

As the number of documents by nationalities is not large enough to achieve statistically significant conclusions only considering the 1,000 most cited papers, the general trends followed by the countries have been considered instead. In this sense, Figure 8 shows the linear correlation between the total number of cites received by a country and the number of articles among the 1,000 most cited of this country. As it can be seen in Figure 8, the correlation index for the lineal fit is high ($R^2=0.923$).

The most clear deviation respect to the average behavior is China. In this case, the number of publications among the 1,000 most cited is lower than expected from the total number of citations received. While China received around half of the United States citations (222,319 vs. 432,155 citations), the number of Chinese articles among the 1,000 most cited is lower than a quarter (75 vs. 315). However, if we compare Europe with United States, they received similar cites (see Part 1 of this study). Different studies have demonstrated both the number of publications and the number of citations received has increased in Chinese publications in the latest years (Zhou and Leydesdorff, 2006), however, Chinese publications are still under represented in the most cited papers. The technical areas are those in which China is closer to United States, especially in physics, chemistry, engineering and emerging areas such as nanoscience and nanotechnology however, the United States has a substantial intrinsic advantage in biomedical research papers (Kostoff, 2008; Fu and Ho, 2013). Furthermore, as it is going to be explained later, a much higher share of international collaborative papers are found in the Chinese highly cited papers than in the average publications.

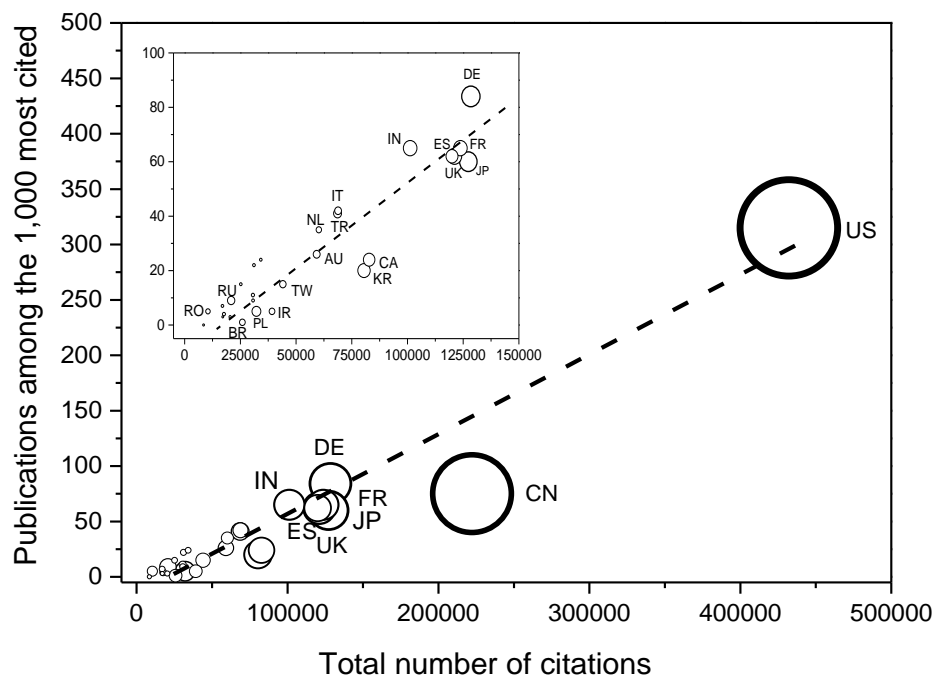

**Figure 8.** Correlation between the total number of cites received by a country and the number of articles among the 1,000 most cited. <u>Notes:</u> Size of the bubbles are proportional to their total production volume in the area.

### 3.5. Degree of international collaboration

International collaboration is usually considered as a way to promote the quality of the Science. In this sense, international collaboration was analyzed for the 1,000 most cited articles and compared with the international collaboration in total scientific production of the area (Part 1), aiming to determine if the highest cited publications have a higher international collaboration degree than the average or not.

Numerous studies showed that internationally co-authored papers attract more citations. In general, the elite structure of top-cited publications is primarily American, but also some European nations increasingly participate in it (Leydesdorff et al., 2014). The other nations participate in this top-layer with a higher probability through international collaborations, especially those countries with a less developed science and technology system. For example, more than 90% of Russian highly cited papers were produced in international collaboration (Pislyakov and Shukshina, 2014), this percentage being more than 80% for Brazilian highly cited papers (Packer and Meneghini, 2006) and a 48% of the Taiwanese most cited articles (Chuang and Ho, 2015). In the case of China, a 47% of their publications included in the ESI database were international collaborative while the average international collaboration for total publications was 17% in SCI (Fu and Ho, 2013). In most cited papers, developed countries usually have lower collaborative ratios than developing ones (Mat et al., 2013). For example, Ho (2012) found for the highly cited articles in Chemical Engineering that the countries with highest percentage of independent articles in total articles were United States (91%), India (89%), Japan (87%), Canada (82%) and Australia (82%), while the countries with the lowest independent articles were China (58%), South Korea (59%), Spain (65%), Greece (66%) or Turkey (68%). Already in 2001, Glänzel (2001) defined different group of countries depending on the effect of the international collaboration on their performance, distinguishing among "big advanced", "small advanced", "big non-advanced" and "small non-advanced" countries.

Among the top 1,000 cited articles, most of the publications (62.9%) are authored by domestic authors, without any international collaboration, thus the international collaboration rate is 37.1%. Furthermore, there are 26.9% of publications with authors from two countries, 6.9% publications with authors from three countries, and a 3.3% publications with authors from four or more countries. There were no important differences in the degree of international collaboration in the 1,000 most cited papers compared to the total scientific production, at least globally. The share of publications with only domestic authors (authors from only one country) was 65.8% for total publications, i.e. the average international collaboration degree was 34.2% (Part 1).

The importance of domestic publications in the most cited papers is slightly lower in Chemical Engineering than in other areas. For example, a 16% of the top 1.1% most cited were independent compared to 16% of internationally collaborative articles (Ho, 2014) and these percentages were similar for the top 1.8% publications in the research area of biomass (24%) (Chen and Ho, 2015), the top 1% cited papers in the area of water resources (30%) (Chuang et al., 2011), or the top 0.68% highly cited papers on health care sciences and service field (13%) (Hsu and Ho, 2014). Nevertheless, if we focus in the top cited papers, i.e. the top 25 highest cited papers in Chemical Engineering, the importance of domestic papers was considerably larger, i.e. 23 of the 25 top cited papers (92%) were domestic papers.

When a possible correlation between the international collaboration degree and the percentage of papers among the 1,000 most cited by countries was analyzed, no correlation was obtained (Figure 9).

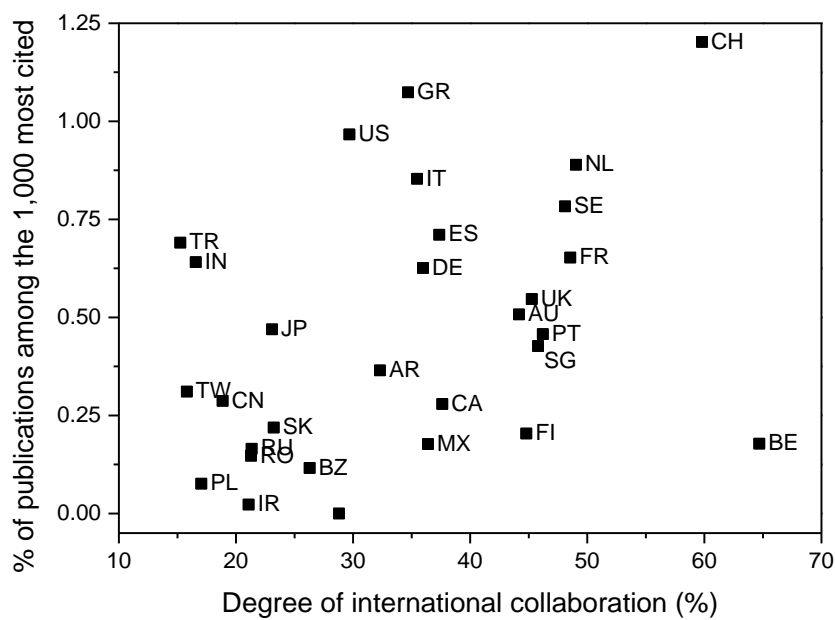

**Figure 9.** Correlation between the percentage of publications among the 1,000 most cited from total publications and the international collaboration degree in total publications by the 30 most productive countries.

The reason is twofold: most cited articles were produced without international collaboration and the rate of international collaboration in the 1,000 most cited articles was very similar to the average total scientific production (37.1% vs. 34.2%). The countries with highest international collaboration in publications, i.e. >40%, had even smaller percentage of papers among the 1,000 most cited papers (average 2.46%) than the countries with low degree of international collaborations, i.e. <25% (average 2.96%). Finally, as commented before, international collaboration was present in only 2 of the 25 most cited papers (8%) in the area.

If we consider EU-27 as a whole, it participated in 336 different publications, distributed as follows: 213 publications performed by a single country (63.4%), 43 publications were carried by two or more European countries (intra-European collaboration) (12.8%) and 80 publications were carried out among European and other countries outside the EU-27 (23.8%). The European Commission stimulates projects among member states and one would expect these R&D programs to lead to more international co-authorship relations in the resulting publications, nevertheless, the member states have retained also a national character, as observed by Leydesdorff et al. (2014), also analyzing highly cited papers.

In the study of Ho (2012), focused in Chemical Engineering, a 90% of the top-cited articles were country independent articles and 10% were internationally collaborative. Furthermore, a 73% of the most cited articles were institution independent and only 27% were inter-institution articles. All the top 10 highly cited papers were published as independent articles. The important difference compared to the present study, in which more than 30% of the highly cited papers were international collaborative, is that the highly cited papers in the study of Ho (2012) were considerably older than in the present study, varying from 1931 to 2010 (only 16% the analyzed documents were published in 2000-2010), while the present study focused on a more recent period (2000-2011). It is well known that the international collaboration is considerably higher in recent papers than in the past due to more exchange of information, information technologies, easier mobility, etc. As stated by Ho (2012), it would be very interesting to analyze why international collaboration is not producing the highest cited papers, in other words, to analyze if international collaboration is producing the best quality papers or just goods papers.

The international collaboration in the 1,000 most cited papers has been also analyzed through a network map using UCINET software (Figure 10). In this map, the nodes represent the different countries and the links represent the collaboration between countries. The size of each link indicates the strength of the international collaboration of a given country while the size of each node indicates total collaborations.

**Figure 10.** International collaborations among the 1,000 most cited publications during the period 2000-2011.

In addition, the centrality degree was calculated. Results of this calculation showed that United States (79), Germany (39), France (32), Spain (31), United Kingdom (28), China (24) and Italy (21) have the highest degree centrality. On the opposite, Portugal (0), Tunisia, Slovakia, Israel and Brazil (1), Belgium, Thailand, Byelarus and Venezuela (2) were those countries with the lowest degree centrality. The highest productive countries have, in general, higher degree centrality, with some exceptions such as Turkey (3), India (8), Canada (9), South Korea (10) and Japan (15).

### 3.6. Journals publishing the most cited articles

The number of journals covered by JCR in the Chemical Engineering subject category varied in the period 2000-2011 from 117 in 2000 to 133 in 2011. Although the total number publications are concentrated in some journals, i.e. the top 10 journals published 29.7% of total scientific production in the area and the first 25 journals, a 50.9%, the most cited articles are even more concentrated. Around half of the most cited articles (48.9%) were published in only 5 journals, 64.6% in 10 journals and 88.6% in 25 journals.

As in Part 1 of this study, the validity of the Bradford's Law for the number of journals publishing the 1,000 most cited papers was tested. In this case, 342 publications were published by the top 3 journals, other 328 publications by other 8 journals and the rest (330 publications) were published in other 52 journals. Therefore, the ratio of journals for publishing each third of the highly cited publications was 3:8:52, equivalent to $1:n:2.4n^2$, which is similar to the results obtained for total publications ($1:n:2.0n^2$) (Part 1) but far from the theoretical Bradfrod's Law ($1:n:n^2$).

Table 2 shows the 15 journals with the highest number of publications among the 1,000 most cited during the 2000-2011 period, representing together around 75% of the 1,000 most cited publications. The journals publishing the highest number of the most cited publications of the area are *Journal of Catalysis, Catalysis Today, Applied Catalysis B – Environment, Journal of Membrane Science and Industrial & Engineering Chemistry Research.* Three of these journals are related specifically to catalysis (all of them published by Elsevier), one to membrane science (also published by Elsevier) and the last is a general research journal in the area (published by the American Chemical Society). If this analysis is extended to the top 15 journals, can be identified: four journals of general research (*Industrial & Engineering Chemistry Research, Chemical Engineering Science, Chemical Engineering Journal and AIChE Journal*), three journals of catalysis (*Journal of Catalysis, Catalysis Today, Applied Catalysis B – Environment*), three journals in energy and fuels (*Progress in Energy and Combustion Science, Fuel, Energy & Fuels*), two journals regarding separation techniques (*Journal of Membrane Science, Separation and Purification Technology*), one journal in dyes and pigments research area (*Dyes and Pigments*), one journal in food science and technology (*Journal of Food Engineering*) and one journal in process biochemistry (*Process Biochemistry*).

In the study of Ho (2012), where the articles with more than 100 citations in the area of Chemical Engineering in the SCI during the period 1931-2010 were analyzed, the top journals were *Journal of Catalysis, AIChE Journal, Chemical Engineering Science and Journal of Membrane Science*. If the journals obtained in the present study are compared with the results obtained by Ho (2012), *Journal of Catalysis* and *Journal of Membrane Science* are among the journals with more top cited articles, while the importance of *Chemical Engineering Science* (ranked 7$^{th}$ in this work) and *AIChE Journal* (ranked 15$^{th}$) is considerably lower. These last two journals are general research journals with a large tradition in Chemical Engineering which have lost some weight in the last years in terms of the number of most cited publications in the area and their impact factor. As recognized by Larivière et al. (2014), since the late 1980s and early 1990s, the quality work has been published in an increasingly diverse and larger group of journals, independently of the area considered, with elite journals publishing a decrease proportion of cited papers, i.e. *Science* and *Nature* journals. It is important to remember that in the study of Ho (2012), an 84% of the considered highly cited publications were published before 2000.

Chuang et al. (2013) using a similar period of time than this study but ESI indicators (1% most cited papers), determined that the journals publishing the highest number of the most cited papers were, in this order: *Energy & Fuels, Journal of Catalysis, Industrial & Engineering Chemistry Research, Combustion and Flame, Catalysis Today, Journal of Membrane Science and Applied Catalysis-B*. Although there are some similarities with the results obtained in the present study, there are also significant differences. The most important difference is that the journal with the highest number of highly cited articles was *Energy & Fuels*, with a 16% of the most cited papers, while this journal was ranked 13[th] (2.3% of the most cited papers) in the present study. In addition, *Combustion and Flame* is ranked 4[th] in the study of Chuang et al. (2013), with an 8.8% of the top cited publications, the same value than this obtained for *Catalysis Today*. However, *Combustion and Flame* is 20[th] ranked in the present study, with only 1.3% of the 1,000 most cited publications.

If the impact factors of the journals are analyzed together with the publications among the most cited in the area, some interesting facts are observed. Theoretically, these values should be highly correlated as the impact factor of a journal is an average of the citations received by the articles published in this journal. An article published in a journal with high impact factor should be more likely to be among the most cited in the area than other published in a journal with a low impact factor. However, Figure 7 demonstrated there is a large variation of the impact factor of the most cited publications.

In this sense, it is interesting to analyze the relationship between the percentage of publications from a journal compared to total scientific production and the percentage of publications of this journal among the 1,000 most cited. Large differences were observed among different journals. For example, despite the high number of publications of *Industrial & Engineering Chemistry Research* (5.8% total publications of the area), this journal published similar and even slightly higher percentage of the most cited papers (6.4%). The same occurs for *Chemical Engineering Science*, *Energy and Fuels, Chemical Engineering Journal*, among others. There is a group of journals in which the percentage of articles among the 1,000 most cited in the area is much higher than the percentage of total publications are: *Journal Membrane of Science* (8.3% vs. 3.0%), *Catalysis Today* (9.6% vs. 2.5%), *Journal of Catalysis* (15.5% vs. 1.9%), *Applied Catalysis B Environmental* (9.1% vs. 1.6%) and *Separation and Purification Technology* (3.1% vs. 1.3%), meaning a very high capacity for publishing highly cited papers. In general, these are the journals with the highest impact factors in the area. On the contrary, there are journals with significantly lower percentage of articles among the most cited in the area compared to its share in total publications, i.e. *Desalination, Chemical Engineering News* or *Powder Technology* (these latest two journals not shown in Table 2), indicating these journals have a low capacity to produce highly cited papers. *Desalination*, ranked 2[th] in terms of total publications (3.3%), was only ranked 28[th] in the 1,000 most cited articles (7 publications, 0.7%); *Chemical Engineering News*, ranked 5[th] in total publications (2.6%), was ranked 37[th] in terms of the most cited papers (3 publications, 0.3%), and

**Table 2.** Top 15 journals publishing the most cited papers in Chemical Engineering during the period 2000-2011.

| Rank in most cited papers | Rank in total publications | Journal | No. papers among the 1,000 most cited | % Top cited papers | No. papers published 2000-2011 | % total papers published | % 1,000 most cited respect total publications | Publisher | Journal country | WoK Areas | Average IF 2000-11 | Median IF 2000-11 |
|---|---|---|---|---|---|---|---|---|---|---|---|---|
| 1 | 11 | Journal of Catalysis | 155 | 15.5 | 4005 | 1.86 | 3.87 | Academic Press Inc. Elsevier Science | United States | Chemistry, Physical Engineering, Chemical | 4.082 | 4.533 |
| 2 | 7 | Catalysis Today | 96 | 9.6 | 5400 | 2.51 | 1.78 | Elsevier Science BV | Netherlands | Chemistry, Applied Chemistry, Physical Engineering, Chemical | 2.696 | 2.696 |
| 3 | 16 | Applied Catalysis B Environmental | 91 | 9.1 | 3382 | 1.57 | 2.69 | Elsevier Science BV | Netherlands | Chemistry, Physical Engineering, Environmental Engineering, Chemical | 4.161 | 3.992 |
| 4 | 4 | Journal of Membrane Science | 83 | 8.3 | 6561 | 3.05 | 1.27 | Elservier Science BV | Netherlands | Engineering, Chemical Polymer Science | 2.342 | 2.270 |
| 5 | 1 | Industrial & Engineering Chemistry Research | 64 | 6.4 | 12477 | 5.80 | 0.51 | American Chemical Society | United States | Engineering, Chemical | 1.614 | 1.511 |
| 6 | 128 | Progress in Energy and Combustion Science | 38 | 3.8 | 215 | 0.1% | 17.67 | Pergamon-Elsevier Science Ltd | United Kingdom | Thermodynamics, Energy & Fuels, Engineering Chemical, Engineering Mechanical | 3.986 | 3.012 |
| 7 | 3 | Chemical Engineering Science | 36 | 3.6 | 6862 | 3.19 | 0.52 | Pergamon-Elsevier Science Ltd. | United States | Engineering, Chemical | 1.794 | 1.770 |
| 8 | 18 | Separation and Purification Technology | 31 | 3.1 | 2876 | 1.34 | 1.08 | Elsevier Science BV | Netherlands | Engineering, Chemical | 1.993 | 2.320 |
| 9 | 17 | Process Biochemistry | 27 | 2.7 | 3060 | 1.42 | 0.88 | Elsevier Sci. Ltd. | United Kingdom | Biochemistry & Molecular Biology, Biotechnology & Applied Microbiology Engineering, Chemical | 1.655 | 1.902 |
| 10 | 12 | Fuel | 25 | 2.5 | 3875 | 1.80 | 0.65 | Elsevier Sci. Ltd. | United Kingdom | Energy & Fuels Engineering, Chemical | 1.664 | 1.521 |
| 11 | 33 | Dyes and Pigments | 24 | 2.4 | 2025 | 0.95 | 1.19 | Elsevier Sci. Ltd. | United Kingdom | Chemistry Applied, Engineering Chemical, Materials Science Textiles | 1.549 | 1.802 |
| 12 | 9 | Chemical Engineering Journal | 23 | 2.3 | 4902 | 2.28 | 0.47 | Elsevier Science SA | Switzerland | Engineering, Environmental Engineering, Chemical | 1.574 | 1.651 |
| 13 | 8 | Energy Fuels | 23 | 2.3 | 5189 | 2.41 | 0.44 | American Chemical Scoiety | United States | Energy & Fuels Engineering, Chemical | 1.706 | 1.507 |
| 14 | 10 | Journal of Food Engineering | 20 | 2.0 | 4330 | 2.01 | 0.46 | Elsevier Sci Ltd. | United Kingdom | Engineering, Chemical Food Science & Technology | 1.440 | 1.473 |
| 15 | 15 | AIChE Journal | 18 | 1.8 | 3394 | 1.58 | 0.53 | Wiley-Blackwell | United States | Engineering, Chemical | 1.699 | 1.777 |

*Powder Technology* was ranked 14$^{th}$ in total publications (1.60%) but only 36$^{th}$ in the most cited papers (10 publications, 1%). Other similar cases were also observed: *Rev. Chemie* (ranked 20$^{th}$ in total publications, 1.33%, and without any article published among the 1,000 most cited), *Korean Journal of Chemical Engineering* (ranked 22$^{th}$ in total publications, 1.0%, and only 1 publication among the 1,000 most cited) and *Chemie Ingenieur Technik* (ranked 24$^{th}$ in total publications, 1.10%, without any article among the 1,000 most cited). In general, these journals have a very low impact factor.

Other interesting aspect to analyze is which are the publishers and their nationalities. The publishers of the top 15 journals publishing the most cited papers are first, Elsevier (8), followed by the Pergamon-Elsevier (2), the American Chemical Society (2), Academic Press Elsevier (1) and Wiley Blackwell (1), similarly to what observed in Part 1 for the top 25 journals in terms of the number of published articles. In the top 15 journals, there are four different journal nationalities: five from the United States, five from the United Kingdom, four from the Netherlands and one from Switzerland. In this case, the importance of United States, United Kingdom and the Netherlands is very similar. However, when the 25 journals with highest scientific production were considered, the number of journals published by United States was considerably higher (9) than those of the Netherlands (6) and United Kingdom (4).

Regarding the Web of Science subject categories of the top 15 journals, four of them were indexed only in Chemical Engineering area (27%), which is a very similar percentage (25%) to that obtained when the top 25 journals in terms of scientific production. However, in the top 25 journals, some of the best ranked journals (1$^{st}$ and 3$^{rd}$) were indexed only in Chemical Engineering subject category, which is not the case for the top 15 journals with higher number of top cited papers, the best positions for these journals are the 5$^{th}$ and the 7$^{th}$. The reason for that is the leading journals in terms of highly cited articles are related to catalysis, which fits well also in other subject categories different to Chemical Engineering such as Physical Chemistry.

### 3.7. Research organizations

Table 3 shows the list of the 35 institutions with the highest number of publications among the 1,000 most cited in the period 2000-2011, i.e. those with seven or more publications. "Organizations-enhanced" search from WoS has been used instead of organizations search to obtain

**Table 3.** Organizations with the highest number of publications among the 1,000 most cited in the area during the period 2000-2011.

| Rank | Organization | Country | No. publications |
|---|---|---|---|
| 1 | Centre National de la Recherche Scientifique (CNRS) | France | 54 |
| 2 | Consejo Superior de Investigaciones Científicas (CSIC) | Spain | 29 |
| 3 | Hong Kong University Science and Technology | China | 19 |
| 4 | Delft University of Technology | Netherlands | 17 |
| 5 | Chinese Academy of Sciences | China | 17 |
| 6 | Indian Institute of Technology | India | 15 |
| 7 | Haldor Topsoe | Denmark | 13 |
| 8 | University of Colorado | United States | 13 |
| 9 | Technical University of Denmark | Denmark | 13 |
| 10 | Swiss Federal Institute of Technology Technol (ETH Zurich) | Switzerland | 13 |
| 11 | Max Planck Institutes | Germany | 13 |
| 12 | Hacettepe University | Turkey | 12 |
| 13 | Penn State University | United States | 12 |
| 14 | University of Minnesota | United States | 11 |
| 15 | University of California Berkeley | United States | 10 |
| 16 | University of Michigan | United States | 10 |
| 17 | Massachusetts Institute of Technology (MIT) | United States | 10 |
| 18 | University of Queensland | Australia | 9 |
| 19 | National University Singapore | Singapore | 8 |
| 20 | Carnegie Mellon University | United States | 8 |
| 21 | Chalmers University of Technology | Sweden | 8 |
| 22 | Texas A&M University | United States | 8 |
| 23 | Toyota | Japan | 8 |
| 24 | University of Twinge | Netherlands | 7 |
| 25 | University of Naples Federico II | Italy | 7 |
| 26 | Exxon Mobil | United States | 7 |
| 27 | French Institute of Petroleum (IFP) | France | 7 |
| 28 | University Laval | Canada | 7 |
| 29 | Karlsruhe Institute of Technology | Germany | 7 |
| 30 | University of Illinois | United States | 7 |
| 31 | University of New South Wales | Australia | 7 |
| 32 | University of Texas | United States | 7 |
| 33 | Sandia National Labs | United States | 7 |
| 34 | University of Patras | Greece | 7 |
| 35 | University of Wisconsin | United States | 7 |

more comprehensive results trying to solve the lack of standardization in the organization names. These 35 research organizations together participated in 414 of the 1,000 most cited publications (41.4%). The main origin of these institutions was United States (13), France (2), China (2),

Netherlands (2), Denmark (2), Australia (2), Germany (2), Spain (1), India (1), Switzerland (1), Greece (1), Canada (1), Australia (1), Italy (1), Singapore (1), Turkey (1) and Japan (1).

If we compare these results with the top 40 institutions in terms of total number of publications (>725 publications) along the period 2000-2011, some comments can be drawn. The 1,000 most cited papers are more concentrated in a smaller number of institutions than the total publications: the top 35 institutions with the highest amount publications among the 1,000 most cited participated in 41.4% of those publications, while the top 40 institutions in total number of publications, participated in only 25.7% of total publications. The number of countries represented in the list of the top organizations is similar: 21 different countries participated among the top 40 organizations in total publications and 17 different countries participated among the top 35 organizations in the 1,000 most cited papers.

There was observed a clear difference between total scientific production (Part 1) and the most cited articles. When the total scientific production in the area was analyzed, the country with more institutions among the 40 most productive was China (8 institutions), followed by United States (5) and France (4). When only the most cited articles were considered, a large difference between United States (13 organizations among the 35 with more cited articles) and the rest of the countries (with two or less institutions among the 35 with more cited articles) was observed. For example, only 2 out of 35 are from China. Furthermore, while the top 5 institutions in total number of publications were from France, China, India, United States and Russia, in the 1,000 most cited papers, the top 5 institutions were from France, Spain, China, Netherlands and China. Although United States had a clear predominance in terms of highly productive institutions, the predominance is less overwhelming than in the past and in other research areas. In the study of H.S. Ho (2012), focused on the highly cited of articles in Chemical Engineering during the period 1931-2010, 8 from the top 10 institutions were in United States (80%). In this case, University of California ranked first, followed by Massachusetts Institute of Technology (MIT) and the University of Texas. Exxon Research and Engineering Company, the only non-university institution in the top, ranked 4[th] in total articles. The only two non-US institutions in the top 10 were Delft University of Technology in Netherlands and University of Tokyo in Japan, ranked 9[th] and 10[th], respectively (Ho, 2012). Similar trends have been obtained in the present study, however, the great achievements of CNRS, CSIC and the Indian Institutes of Technology could be masked in the analysis of Ho (2012), as the search "organizations-enhanced" provided by WoS was not available at the moment of the analysis carried out by H.S. Ho (2012). These large organizations are those precisely pretended to be more efficiently represented through this enhanced search tool. In other study, 14 from the top 15 institutions publishing the highest cited review papers in SCI expanded (1899-2011) were from the United States (Ho and Kahn, 2014). Furthermore, 16 of the top 22 institutions publishing the highly cited papers in materials science (1900-2010) (Ho, 2014) and 16 of the top 21 institutions publishing the highly cited papers in biomass research (1946-2011) (Chen and Ho, 2015) were from the United States.

Although United States had the highest number of institutions publishing highly cited papers, the two institutions with the highest number of publications among the 1,000 most cited are from France (CNRS) and Spain (CSIC). CNRS (the French National Center for Scientific Research) is formed by 10 institutes, with over 32,000 staff and €3,400 million annual budget. Among CNRS, the most productive institutes were the Univ. Paris 06 (8 publications), the Institut des Recherches sur la Catalyse et l'Environnement de Lyon (8 publications), the École Central de Paris (6 publications) and the École Central de Lyon (5 publications). On the other hand, CSIC (the Spanish Scientific Research Council) is formed by 125 centers and institutes, with 12,795 staff and €731 million annual budget. The most productive institutes from CSIC are the Instituto Nacional del Carbón y el Instituto Carboquímica (11 publications) and the Instituto Tecnología Química (7 publications). CNRS was also ranked 1$^{st}$ in terms of total scientific production (2.12% world production), however, CSIC was ranked only the 9$^{th}$ (0.80% world production). Despite 13 of top 35 institutions publishing highly cited papers are from United States, the first ranked institution was the University of Colorado (8$^{th}$ ranked),

There was also remarkable the position of the Hong Kong University of Science and Technology (China) (ranked 3$^{rd}$, 19 publications) and Delft University of Technology (Netherlands) (ranked 4$^{th}$, 15 publications), especially considering they are not large universities. In the case of Hong Kong University of Science and Technology, it has around 13,079 students, with only 627 teaching staff, and research funding of $708 million (data from 2013). In the University of Delft, there are 8 faculties and 3 research institutes, it has around 18,781 students with 2,579 academic staff, and around €900 million budget (data also from 2013).

It was also interesting to notice the presence of three companies among the top 35 research organizations publishing the highest cited papers: Haldor Topsoe from Denmark (13 publications), Toyota from Japan (8 publications) and Exxon Mobil from United States (7 publications). This was not observed when total scientific production were analyzed: none of 40 most productive organizations were companies.

## 4. CONCLUSIONS

- United States, Europe and Asia have similar shares in the 1,000 most cited papers of the area, however, there are important differences: United States is overrepresented compared to its share in total publications (31.5% vs. 15.3%), Europe is represented about to expected by their share in total publications (33.7% vs. 34.6%) and Asia is underrepresented (31.3% vs. 40.5%).

- Although there is a large predominance of the United States in the most cited papers of the area, there are some small countries with higher percentage of published articles among the most cited in the area such as Denmark (1.68%), Switzerland (1.20%), Greece (1.07%) or Ireland (0.97%), compared to the United States (0.95%).

- A displacement of the nationalities of the authors of the 1,000 most cited papers to the Far East has been observed, but in a much lower extent than the total scientific production in the area (Part 1). Although China improved its position, United States maintains its outstanding position. India and Iran are less important in terms of the most cited papers than in total scientific production, and Turkey improved largely its importance in producing highly cited papers in the last years.

- The most cited papers are published in journals with higher impact factor than the average, however, there is a significant number of articles among the 1,000 most cited papers in journals with low impact factors, i.e. lower than 1.

- The articles among the 1,000 most cited correlates well with the number of total citations received by countries, however China share in the most cited publications is still lower than expected by the total citations received.

- The international collaboration is very similar in total publications and in the 1,000 most cited papers of the area, i.e. 34.2% vs. 37.1%, respectively. It seems there is not a clear relationship between the 1,000 most cited papers and the degree of international collaboration. In fact, of the 25 most cited papers in the period 2000-2011, only 2 were international collaborations (8%).

- The 1,000 most cited papers are concentrated in a few number of journals. Around half of the most cited papers were published in only five journals: *Journal of Catalysis, Catalysis Today, Applied Catalysis B-Environment, Journal of Membrane Science and Industrial & Engineering Chemistry Research*.
- Although the highest number of institutions with publications among the most cited in the area are from United States (13 from 35), the two institutions with the highest number of publications among the 1,000 most cited are CNRS (France) and CSIC (Spain).